\RequirePackage{tikz}

\documentclass[doublecol]{epl2}

\usepackage{graphicx}
\usepackage{xcolor}
\usepackage{tikz}
\usepackage[caption=false]{subfig}
\usepackage{mathrsfs}
\usepackage{
amssymb}
\usepackage{mathtools}
\usepackage{physics}
\usepackage{bm}
\usepackage{listings}
\usepackage{cases}
\usepackage{comment}
\usepackage{soul}
\usepackage{cancel}
\usepackage{cases}
\usepackage[utf8]{inputenc}
\usepackage{url}
\usepackage{longtable}
\usepackage[normalem]{ulem}
\usepackage{xspace}
\usepackage[colorlinks=true
,urlcolor=darkblue
,anchorcolor=darkblue
,citecolor=darkblue
,filecolor=darkblue
,linkcolor=darkblue
,menucolor=darkblue
,linktocpage=true
,pdfproducer=medialab
,pdfa=true
]{hyperref}

\newcommand{\ee}{\mathrm{e}}
\newcommand{\diag}{\mathrm{diag}}
\newcommand{\Mpl}{M_\text{Pl}}
\newcommand{\ns}{n_{{}_\mathrm{S}}}

\newcommand{\SO}{\mathrm{SO}}
\newcommand{\groupO}{\mathrm{O}}

\newcommand{\calL}{\mathcal{L}}

\newcommand{\calP}{\mathcal{P}}

\newcommand{\bae}[1]{\begin{align} #1 \end{align}}

\newcommand{\dps}{\displaystyle}

\definecolor{monza}{HTML}{CF000F}
\definecolor{darkblue}{HTML}{00008b}
\definecolor{darkmagenta}{HTML}{8b008b}

\title{Conformal inflation in the metric-affine geometry}
\date{\today}

\author{Y. Mikura\inst{1}\thanks{E-mail: \email{mikura.yusuke@e.mbox.nagoya-u.ac.jp}} \and 
Y. Tada\inst{1}\thanks{E-mail: \email{tada.yuichiro@e.mbox.nagoya-u.ac.jp}} 
\and 
S. Yokoyama\inst{2,3}\thanks{E-mail: \email{shu@kmi.nagoya-u.ac.jp}}
}

\institute{
    \inst{1} Department of Physics, Nagoya University, Nagoya 464-8602, Japan
    
    \inst{2} Kobayashi Maskawa Institute, Nagoya University, Chikusa, Aichi
    464-8602, Japan
    
    \inst{3} Kavli IPMU (WPI), UTIAS, The University of Tokyo, Kashiwa,
    Chiba 277-8583, Japan
}


\abstract{
Systematic understanding for classes of inflationary models is investigated from the viewpoint of the local conformal symmetry and the slightly broken global symmetry in the framework of the metric-affine geometry. In the metric-affine geometry, which is a generalisation of the Riemannian one adopted in the ordinary General Relativity, the affine connection is an independent variable of the metric rather than given e.g. by the Levi-Civita connection as its function. Thanks to this independency, the metric-affine geometry can preserve the local conformal symmetry in each term of the Lagrangian contrary to the Riemannian geometry, and then the local conformal invariance can be compatible with much more kinds of global symmetries. As simple examples, we consider the two-scalar models with the broken $\SO(1,1)$ or $\groupO(2)$, leading to the well-known $\alpha$-attractor or 
natural inflation, respectively.
The inflaton can be understood as their pseudo Nambu-Goldstone boson.
}

\begin{document}

\maketitle

\section{Introduction}

Since the advent of the concept of cosmic inflation, hundreds of theoretical models have been studied for its realisation, though a unanimous understanding is yet to be obtained. 
Amongst these vast numbers of inflationary models, recent remarkable progress of cosmological observations represented by the precise measurement of the cosmic microwave background (CMB) by the Planck Collaboration has been revealing the favoured classes of inflation~\cite{Akrami:2018odb}. 
The first attempt to inflation by
Starobinsky~\cite{Starobinsky:1980te} represents such favoured classes. A possible generalisation of this class has recently been called \emph{$\alpha$-attractor}~\cite{Kallosh:2013lkr, Kallosh:2013yoa,Kallosh:2014rga}, in which the effective potential for the canonical scalar (inflaton) $\varphi$ has a form of
\bae{
    V(\varphi)\sim\Lambda^4\tanh^{2n}\pqty{\frac{\varphi}{\sqrt{6\alpha}}}, \qquad n=1,2,3,\cdots.
}
This class of inflationary models uniformly predicts
the scalar spectral index $\ns$ and the tensor-to-scalar ratio $r$ as~\cite{Kallosh:2013yoa}
\bae{
    \ns\coloneqq\dv{\log\calP_\zeta}{\log k}\simeq 1-\frac{2}{N}~, \qquad r\coloneqq\frac{\calP_h}{\calP_\zeta}\simeq \frac{12\alpha}{N^2}~,
}    
independently of the power $n$ for small $\alpha$. Here $\calP_\zeta$ and $\calP_h$ denote the primordial scalar and tensor power spectra, respectively, and $N$ is the number of e-folds of inflation. 
They are well consistent with the Planck 2018's $2\sigma$ constraints $\ns|_{k=0.05\,\mathrm{Mpc}^{-1}}=0.9649\pm0.0084$ and $r|_{k=0.002\,\mathrm{Mpc}^{-1}}<0.056$~\cite{Akrami:2018odb}
for $N\sim 50\text{--}60$ and $\alpha\lesssim10$ in Planck mass unit (Starobinsky's model corresponds with $\alpha=1$). Another possibility, natural inflation~\cite{Freese:1990rb}, where the potential is given by the periodic form
\bae{
    V(\varphi)\sim\Lambda^4\pqty{1-\cos\frac{\varphi}{f}},
}
is also marginally consistent with the Planck's constraint if $f\sim7$.

The further advantage of the $\alpha$-attractor is its compatibility with the local conformal symmetry. The conformal or scale invariance is an important concept in many physical contexts including cosmology (see, e.g., Ref.~\cite{Bars:2013yba}). We have however found neither the scale invariance nor the corresponding massless Nambu-Goldstone (NG) mode at low energy, so it must be broken explicitly or implemented as a local symmetry. 
The local conformal symmetry is also helpful for, e.g., supergravity embedding~\cite{Kallosh:2000ve,Ferrara:2010in} (see also, e.g., Refs. \cite{Buchmuller:2012ex,Choudhury:2013zna,Buchmuller:2013zfa,Ishiwata:2018dxg}).
Inflation with the local conformal symmetry called \emph{conformal inflation} has been studied well (see, e.g., Refs.~\cite{Kallosh:2013pby,Kallosh:2013hoa}).
The $\alpha$-attractor in this context can be understood as the pseudo-NG mode of the additional global symmetry with a small explicit breaking, so that the flatness of inflaton’s potential is protected.
However the local conformal invariance restricts the relation between the scalar kinetic term and its coupling to the Ricci scalar in the ordinary Riemannian geometry and thus the idea of conformal inflation cannot be freely generalised to other global symmetry groups.
In fact natural inflation, which is also interpreted as the pseudo-NG mode, cannot be implemented in a local-conformal way (see Ref.~\cite{Tang:2019olx} for an attempt to implement natural inflation in the local-conformal action by introducing the dynamical Weyl gauge field.)

In this 
Letter we show that these two classes of inflation: $\alpha$-attractor and natural inflation can be systematically understood by the local conformal symmetry and the slightly broken global symmetry in the framework of the metric-affine geometry where the metric and affine connection are treated as independent variables. This
generalised
geometry implies a possibility that many kinds of inflation could be further unified or some novel class of inflation could be developed in the context of the local conformal invariance. Throughout this paper, we adopt the Planck unit $c=\hbar=\Mpl=1$ and the sign of the Minkowski metric is defined by $\eta_{\mu\nu}=\diag(-1,1,1,1)$.

\section{
The metric-affine geometry and the local conformal transformation}
The usual General Relativity employs the (pseudo-)Riemannian geometry, where only the metric $g$ is an independent variable and the affine connection $\Gamma$ is its dependent function, 
given e.g., by the Levi-Civita connection:
\bae{
    \Gamma^\rho_{\mu\nu}(g)=\frac{1}{2}g^{\rho\lambda}\pqty{\partial_\mu g_{\nu\lambda}+\partial_\nu g_{\mu\lambda}-\partial_\lambda g_{\mu\nu}}.
}
The
local conformal transformation in this geometry is defined as the change of metric by a scalar factor at each spacetime point as\footnote{Strictly speaking, this metric transformation should be called the \emph{Weyl transformation}, while the \emph{conformal transformation} means the change of the coordinate. Nevertheless we refer to this metric transformation as the \emph{conformal transformation} in this work, following the convention of the community.}
\bae{
    g_{\mu\nu}\to\tilde{g}_{\mu\nu}=\ee^{-2\sigma(x)}g_{\mu\nu},
}
accompanied by the corresponding transformation of the Levi-Civita connection. Due to this non-trivial transformation of the connection,
the Ricci scalar is not covariant under this transformation as can be seen from
\bae{
    R(g)\to\tilde{R}(g)=\ee^{2\sigma(x)}\pqty{R(g)-6\ee^{\sigma(x)}\Box\ee^{-\sigma(x)}},
\label{eq:Ricci_trans}
}
in the 4-dimensional spacetime.
One thus often introduces another transforming scalar field $S$,
\bae{
    g_{\mu\nu}\to\tilde{g}_{\mu\nu}=\ee^{-2\sigma(x)}g_{\mu\nu}, \qquad S\to\tilde{S}=\ee^{\sigma(x)}S,
}
to 
construct a local-conformal-invariant Lagrangian
\bae{
    \calL\supset\sqrt{-g}\pqty{\frac{1}{12}S^2R(g)+\frac{1}{2}g^{\mu\nu}\partial_\mu S\partial_\nu S}.
    \label{eq:Lag_conformal}
}
The extra $\sigma$-derivatives in eq.~\eqref{eq:Ricci_trans} can cancel thanks to the kinetic term of the scalar field.
The extra scalar factor $\ee^{-2\sigma}$ in $\sqrt{-g}R$ is also cancelled by the antifactor from $S^2$ in the non-minimal coupling term. One may also add the quartic potential $-\sqrt{-g}\frac{1}{4}\lambda S^4$ which has also local conformal invariance by itself.

Now the 
Einstein gravity can be understood as a particular gauge choice $S=\sqrt{6}$ of the local conformal symmetry.
It could be also seen as one variety of the \emph{induced gravity} scenario~\cite{Zee:1978wi,Accetta:1985du}.
The scalar $S$ appears inevitably as a ghost in this context because the
proportion of the coefficient of each term in eq.~\eqref{eq:Lag_conformal} should be specified to cancel the $\sigma$-derivatives and then the kinetic term of $S$ exhibits the wrong sign.
This is not problematic in itself as $S$ is a ``fixed" degree of freedom (DoF) and removed from the theory. However then one cannot have any matter component other than gravity in this minimal setup. 
One may add another scalar DoF as an inflaton to the theory
but the coefficients of their non-minimal couplings and kinetic terms are still restrictive.

\bigskip

On the other hand one can generalise the geometry side to the so-called \emph{metric-affine} one where both the metric and the affine connection are treated as independent variables (see, e.g., \cite{Hehl:1994ue,Kleyn:2004yj,Vitagliano:2010sr, Vitagliano:2013rna,Vazirian:2013baa,Shimada:2018lnm}
and references therein).
If the gravity part is dictated only by the Einstein-Hilbert action, the Lagrangian constraint restricts the connection to the 
Levi-Civita one with the torsion-free condition as a gauge choice, and the two geometries do not lead to any difference. 
In general, the metric-affine geometry however exhibits different physics from the Riemannian even if the action takes the same form.

In the metric-affine geometry,  
the connection is left unaffected under the conformal transformation as an independent DoF 
of the metric:
\bae{
    g_{\mu\nu}\to\tilde{g}_{\mu\nu}=\ee^{-2\sigma(x)}g_{\mu\nu}, \qquad \Gamma^\mu{}_{\alpha\beta}\to\tilde{\Gamma}^\mu{}_{\alpha\beta}=\Gamma^\mu{}_{\alpha\beta}.
}
As the Riemann tensor is a function only of the connection, it obviously leads to the covariant Ricci scalar:
\bae{
    R(g,\Gamma)=g^{\mu\nu}R_{\mu\nu}(\Gamma)\to R(\tilde{g},\tilde{\Gamma})=\ee^{2\sigma(x)}R(g,\Gamma).
}
Its large benefit is that the non-minimal coupling term $\sqrt{-g}S^2R(g,\Gamma)$ exhibits the local conformal invariance by itself without specifying the kinetic term of the scalar $S$.
The conformal invariance of the kinetic term can be also restored in itself by replacing the ordinary derivatives by the covariant ones defined by~\cite{2019Univ....5...82I}
\bae{
    D_\mu\coloneqq\partial_\mu-\frac{1}{8}Q_\mu,
}
with the non-metricity~\cite{Kleyn:2004yj}
\bae{
    Q_\mu=g^{\alpha\beta}Q_{\mu\alpha\beta}\coloneqq-g^{\alpha\beta}\nabla_\mu g_{\alpha\beta}, 
}
which vanishes for a metric-compatible connection.
One then finds that this derivative transforms covariantly as
\bae{
    D_\mu S\to\tilde{D}_\mu\tilde{S}=\ee^{\sigma(x)}D_\mu S.
}
To put it short, in the metric-affine geometry, each of the non-minimal coupling term $\sqrt{-g}S^2R(g,\Gamma)$, the scalar kinetic term $\sqrt{-g}g^{\mu\nu}D_\mu SD_\nu S$, and the possible potential term $\sqrt{-g}S^4$ has independently local conformal invariance. 

\bigskip

To clarify the role of the non-metricity $Q_\mu$, the connection transformation called \emph{projective transformation} peculiar to the metric-affine geometry~\cite{Sotiriou:2006qn,2019Univ....5...82I} 
\bae{
    \Gamma^\mu{}_{\alpha\beta}\to\tilde{\Gamma}^\mu{}_{\alpha\beta}=\Gamma^\mu{}_{\alpha\beta}+\delta^\mu_\alpha\xi_\beta(x),
}
is useful.
It can be easily proven that the Ricci scalar is invariant as $\tilde{R}(g,\tilde{\Gamma})=R(g,\Gamma)$ under the projective transformation, while 
the non-metricity 
changes as 
\begin{equation}
    Q_\mu\to\tilde{Q}_\mu=Q_\mu+8\xi_\mu.
\end{equation}
Without any explicit $Q_\mu$-term  
in the Lagrangian, 
the action thus enjoys the local 
projective invariance and the non-metricity is a mere gauge choice.
It ensures that one can arbitrarily choose $Q_\mu$ and adopt, e.g., the metric-compatible connection: $Q_\mu=0$. If the action is constituted only by the Einstein-Hilbert one with minimally coupled matters, the stationary constraint on the connection with this gauge choice leads to the ordinary Levi-Civita connection and the metric-affine geometry coincides with the Riemann formulation \cite{Dadhich:2010xa}.
On the other hand,  
the explicit $Q_\mu$-term  
breaks the local projective invariance and $Q_\mu$ is recognised as a physical (but non-dynamical) DoF.  
Nonetheless, since the Ricci scalar (and trivially the potential term) 
has the projective invariance, it does not influence the $Q_\mu$'s 
stationary solution and the non-metricity can be integrated out only through the  
conformal kinetic term.
We refer the reader to Table~\ref{table: trs. list} for a list of the transformation laws both in the Riemannian and metric-affine geometries as a summary of this section.

\begin{largetable}
    \centering
    \begin{tabular}{|c|c|c|}
        \hline
         & \textbf{Riemannian} & \textbf{Metric-affine} \\
        \hline
        \parbox{
            0.25\hsize}{
            \textbf{Conformal trans.}
            \begin{align*}
                \begin{array}{lll}
                    \dps
                    g_{\mu\nu} &
                    \dps
                    \!\!\to\tilde{g}_{\mu\nu} & 
                    \dps
                    \!\!=e^{-2\sigma(x)}g_{\mu\nu} \\
                    \dps
                    S & 
                    \dps
                    \!\!\to\tilde{S} &
                    \dps
                    \!\!=e^{\sigma(x)}S
                \end{array}
            \end{align*}} &
        \parbox{
            0.4\hsize}{
            \begin{align*}
                \begin{array}{ll}
                    \dps
                    \tilde{\Gamma}^\rho_{\mu\nu} & 
                    \dps
                    =\Gamma^\rho_{\mu\nu}-\delta^\rho_\mu\partial_\nu\sigma-\delta^\rho_\nu\partial_\mu\sigma+g_{\mu\nu}g^{\rho\lambda}\partial_\lambda\sigma \\
                    \dps
                    \tilde{R} & 
                    \dps
                    =e^{2\sigma}(R-6e^\sigma\Box e^{-\sigma}) \\
                    \dps
                    \partial_\mu\tilde{S} & 
                    \dps
                    =e^{\sigma}(\partial_\mu S+S\partial_\mu \sigma)
                \end{array}
            \end{align*}} &
        \parbox{
            0.3\hsize}{
            \begin{align*}
                \begin{array}{ll}
                    \dps
                    \tilde{\Gamma}^\rho_{\mu\nu} &
                    \dps
                    =\Gamma^\rho_{\mu\nu} \\
                    \tilde{R} & 
                    \dps
                    =e^{2\sigma}R \\
                    \dps
                    \tilde{Q}_\mu & 
                    \dps
                    =Q_\mu+8\partial_\mu\sigma \\
                    \dps
                    \tilde{D}_\mu\tilde{S} & 
                    \dps
                    =e^{\sigma}D_\mu S
                \end{array}
            \end{align*}
        }
        \\ \hline
        \parbox{
            0.25\hsize}{
            \vspace{10pt}\textbf{Projective trans.} 
            \begin{align*}
                \Gamma^\mu_{\alpha\beta}\to\tilde{\Gamma}^\mu_{\alpha\beta}=\Gamma^\mu_{\alpha\beta}+\delta^\mu_\alpha\xi_\beta
            \end{align*}} &
        \tikz[baseline=0pt]{
            \useasboundingbox(0,0);
            \draw(-3.8,1)--(3.8,-0.85);} &
        \parbox{
            0.3\hsize}{
            \begin{align*}
                \begin{array}{ll}
                    \dps
                    \tilde{R} &
                    \dps
                    =R \\
                    \dps
                    \tilde{Q}_\mu & 
                    \dps
                    =Q_\mu+8\xi_\mu
                \end{array}
            \end{align*}
        }
        \\\hline
    \end{tabular}
    \caption{The transformation laws of variables in the ordinary Riemannian geometry and the metric-affine geometry.}
    \label{table: trs. list}
\end{largetable}

\section{
Conformal inflation}

In the previous section, we saw that, in the metric-affine geometry, an arbitrary combination of the non-minimal coupling $\sqrt{-g}S^2R(g,\Gamma)$, the scalar kinetic term $\sqrt{-g}D_\mu SD^\mu S$, and its potential $\sqrt{-g}S^4$ exhibits the local conformal invariance. However this scalar DoF $S$ could be removed by the gauge fixing of the conformal symmetry as, e.g.,  $S=\mathrm{const.}$, where the action is reduced to the mere Einstein-Hilbert one with a cosmological constant after integrating out $Q_\mu$. 
One minimal extension is thus adding another scalar to preserve one inflaton DoF as we investigate below.

\subsection{
Global symmetry}

We further impose an additional global symmetry on the two scalar fields $\phi$ and $\chi$ to keep the inflaton's effective potential flat. As a first example, let us
start with
the most general Lagrangian which respects  
both the local conformal 
and the global $\SO(1,1)$ symmetry:
\bae{
    \mathcal{L}=\sqrt{-g}\left[\frac{1}{12\alpha}(\chi^2-\phi^2)R(g,\Gamma)+\frac{1}{2}D_\mu\chi D^\mu\chi \right. \nonumber \\
    \left.-\frac{1}{2}D_\mu\phi D^\mu\phi-\frac{1}{4}\lambda(\chi^2-\phi^2)^2 \right]\label{lambda},
}
with arbitrary coupling constants $\alpha$ and $\lambda$ (the particular notation of the coefficient $1/12\alpha$ is for later convenience).
We note that, contrary to the ordinary Riemannian geometry, $\alpha$ is not necessarily fixed to unity as each term independently respects the local conformal invariance in the metric-affine geometry.
Making use of the local conformal symmetry, it can be simplified by fixing the scalars to, e.g., $\chi^2-\phi^2=6\alpha$ called \emph{rapidity gauge}~\cite{Kallosh:2013hoa,Kallosh:2013yoa}. This unifies the two scalars to one canonical field $\varphi$ through
\begin{equation}
    \label{eq: rapidity}
    \chi=\sqrt{6\alpha}\cosh\frac{\varphi}{\sqrt{6\alpha}}, \qquad \phi=\sqrt{6\alpha}\sinh\frac{\varphi}{\sqrt{6\alpha}},
\end{equation}
and the model we consider is expressed as
\begin{equation}
    \mathcal{L}=\sqrt{-g}\left[\frac{1}{2}R(g,\Gamma)-\frac{1}{2}\partial_\mu\varphi\partial^\mu\varphi+\frac{3\alpha}{64}Q_\mu Q^\mu-
    9\lambda \alpha^2\right].
\end{equation}
$Q_\mu$'s stationary solution is trivial as $Q_\mu=0$ and 
one obtains 
the Lagrangian for a free massless scalar $\varphi$ with a cosmological constant $\Lambda^4\coloneqq 
9\lambda\alpha^2$ as 
\begin{equation}
    \mathcal{L}=\sqrt{-g}\left[\frac{1}{2}R(g,\Gamma|_{Q_\mu=0})-\frac{1}{2}\partial_\mu\varphi\partial^\mu\varphi-\Lambda^4\right].
\end{equation}

It can be also checked with another gauge choice, e.g., the \emph{conformal gauge} $\chi=\sqrt{6\alpha}$. In this gauge, the Lagrangian first reads
\bae{
    \mathcal{L}=\sqrt{-g}\left[\frac{1}{2}\left(1-\frac{\phi^2}{6\alpha}\right)R(g,\Gamma) 
    +\frac{3\alpha}{64}Q_\mu Q^\mu \right. \nonumber \\
    \left.
   -\frac{1}{2}D_\mu\phi D^\mu\phi
    -9\lambda\alpha^2\left(1-\frac{\phi^2}{6\alpha}\right)^2 \right].
}
This so-called Jordan frame expression can be simplified to the Einstein frame by the conformal redefinition of the metric: $g_{\mu\nu}\to\pqty{1-\frac{\phi^2}{6\alpha}}g_{\mu\nu}$ 
(but leaving other variables
including the connection $\Gamma$ unaffected in the metric-affine geometry) as
\bae{
    \mathcal{L}&=\sqrt{-g}\left[\frac{1}{2}R(g,\Gamma)+\frac{3\alpha}{64}Q_\mu Q^\mu \right. \nonumber \\
    &\qquad\qquad\qquad\left.-\frac{1}{2}\frac{1}{(1-\phi^2/6\alpha)^2}\partial_\mu \phi\partial^\mu\phi-\Lambda^4\right].
}
Taking the constraint $Q_\mu=0$ into account, this Lagrangian again boils down to the canonical free scalar $\varphi\coloneqq\sqrt{6\alpha}\tanh^{-1}\frac{\phi}{\sqrt{6\alpha}}$ with a cosmological constant $\Lambda^4=9\lambda\alpha^2$. This is not a characteristic feature only of the $\SO(1,1)$ symmetry. In the metric-affine geometry, one can instead impose, e.g., the global $\groupO(2)$ symmetry as
\bae{
    \calL=\sqrt{-g}\left[\frac{1}{8f^2}(\phi^2+\chi^2)R(g,\Gamma)-\frac{1}{2}D_\mu\phi D^\mu\phi \right. \nonumber \\
    \left.-\frac{1}{2}D_\mu\chi D^\mu\chi-\frac{1}{4}\lambda(\phi^2+\chi^2)^2\right],
}
with arbitrary parameters $f$ and $\lambda$ (again the notation $1/8f^2$ is for later convenience).
Taking the \emph{rotational
gauge} $\phi^2+\chi^2=4f^2$ by
\bae{\label{eq: rotational}
    \phi=2f\cos\frac{\varphi}{2f}, \qquad \chi=2f\sin\frac{\varphi}{2f},
}
one then easily finds that it also gives the free massless scalar with a cosmological constant.

We saw that both the exact $\SO(1,1)$ and $\groupO(2)$ symmetry equally 
lead to the free massless scalar with a cosmological constant.
However once they are explicitly broken, they give rise to two different inflationary models both of which are well motivated by the CMB observation, as we will see in the next subsection.

\subsection{
Breaking the global 
symmetry}

To make the inflaton potential slightly tilted, let us introduce a small explicit breaking to the global symmetry, keeping the local conformal symmetry. First the broken $\SO(1,1)$ model can be given, e.g., by
\bae{
    \mathcal{L}=\sqrt{-g}\left[\frac{1}{12\alpha}(\chi^2-\phi^2)R(g,\Gamma)+\frac{1}{2}D_\mu\chi D^\mu\chi \right. \nonumber \\
    \left.-\frac{1}{2}D_\mu\phi D^\mu\phi-\frac{1}{36\alpha^2}F\left(\frac{\phi}{\chi}\right)(\chi^2-\phi^2)^2 \right], \label{alpha}
}
where $F$ is an arbitrary function and a combination $\phi/\chi$ is a unique way of preserving the local conformal symmetry~\cite{Kallosh:2013hoa}. Its coefficient $1/36\alpha^2$ is just for later convenience. 
As this model still has
the local conformal
invariance, one can again fix the scalar fields as a gauge choice. 
In the rapidity gauge $\chi^2-\phi^2=6\alpha$~\eqref{eq: rapidity}, 
it reads
\bae{
    \mathcal{L}=\sqrt{-g}\left[\frac{1}{2}R(g,\Gamma)-\frac{1}{2}\partial_\mu\varphi\partial^\mu\varphi+\frac{3\alpha}{64}Q_\mu Q^\mu \right. \nonumber \\
    \left.-F\left(\mathrm{tanh}\frac{\varphi}{\sqrt{6\alpha}}\right)\right],
}
and
the stationary solution $Q_\mu=0$  
eventually leads to 
\begin{equation}
    \mathcal{L}\!=\!\sqrt{-g}\left[\frac{1}{2}R(g,\Gamma|_{Q_\mu=0})\!-\!\frac{1}{2}\partial_\mu\varphi\partial^\mu\varphi\!-\!F\left(\mathrm{tanh}\frac{\varphi}{\sqrt{6\alpha}}\right)\right].
\end{equation}
It is nothing but the well-known $\alpha$-attractor inflation with the monomial potential 
$F(x)=x^{2n}$ ($n=1,2,3,\cdots$)~\cite{Kallosh:2013yoa}.\footnote{Higgs inflation~\cite{Bezrukov:2007ep}
which corresponds to $\phi^4$-chaotic inflation model with non-minimal coupling to gravity (see Ref.~\cite{Kallosh:2013tua}
for its generalisation)
is well known to give very similar observational predictions
to Starobinsky's model in the Riemannian geometry.
Interestingly,
such a Higgs-like inflation
in the metric-affine geometry/Palatini formalism
can be almost equivalent to the $\alpha$-attractor model~\cite{Bauer:2008zj,Aoki:2018lwx,Takahashi:2018brt}
(see Ref.~\cite{Jarv:2017azx} for its generalisation).} 
Intriguingly in the metric-affine geometry, the $\alpha$ parameter can be easily (and inevitably in a general Lagrangian) introduced as a coupling constant of the non-minimal coupling, thanks to the feature that the local conformal invariance can hold in each term independently.
In the conformal gauge $\chi=\sqrt{6\alpha}$, one can also see another aspect: pole inflation~\cite{Galante:2014ifa,Broy:2015qna}
as
\bae{
    \calL\!=\!\sqrt{-g}\left[\frac{1}{2}R(g,\Gamma|_{Q_\mu=0})\!-\!\frac{1}{2}\frac{\partial_\mu\phi\partial^\mu\phi}{(1-\phi^2/6\alpha)^2}\!-\!F\pqty{\frac{\phi}{\sqrt{6\alpha}}}\right].
}

As well as $\alpha$-attractor inflation, one can realise other inflationary models in a similar framework because there is no longer restrictive relation between the non-minimal couplings and the scalar kinetic terms in the metric-affine geometry. If one considers the broken $\groupO(2)$ model as
\bae{
    \calL=\sqrt{-g}\left[\frac{1}{8f^2}(\phi^2+\chi^2)R(g,\Gamma)-\frac{1}{2}D_\mu\phi D^\mu\phi \right. \nonumber \\
    \left.-\frac{1}{2}D_\mu\chi D^\mu\chi-\frac{1}{16f^4}F
   \left(\frac{\phi}{\chi}\right)(\phi^2+\chi^2)^2\right],
}
the rotational gauge~\eqref{eq: rotational} leads to
\bae{
    \calL=\sqrt{-g}\left[\frac{1}{2}R(g,\Gamma)-\frac{1}{2}\partial_\mu\varphi\partial^\mu\varphi-\frac{f^2}{32}Q_\mu Q^\mu \right. \nonumber \\
    \left.-F\left(1\Bigm/\tan\frac{\varphi}{2f}\right)\right].
}
For, e.g., $F(x)=2\Lambda^4/(1+x^2)$ (see footnote\footnote{Though its form seems artificial, its boundedness is necessary for the ``small" explicit breaking of $\groupO(2)$, contrary to the $\alpha$-attractor case where $\phi=\sqrt{6\alpha}\tanh\frac{\varphi}{\sqrt{6\alpha}}$ itself is already bounded. In fact it is not unnatural as it reads the ordinary renormalisable potential $V(\phi,\chi)\propto\chi^2(\phi^2+\chi^2)$ in terms of $\phi$ and $\chi$.}),
one obtains 
the natural inflation in this case~\cite{Freese:1990rb}:
\bae{
    \calL\!=\!\sqrt{-g}\left[\frac{1}{2}R(g,\Gamma|_{Q_\mu=0})\!-\!\frac{1}{2}\partial_\mu\varphi\partial^\mu\varphi\!-\!
    \Lambda^4\!\left(1-\cos\frac{\varphi}{f}\right)\right].
}
Contrary to the axion-type natural inflation, the ``decay" constant $f$ need not be smaller than the Planck scale and thus it can be easily compatible with the CMB observation, that is, one can naturally take $f \sim 7$.

\section{Conclusions}

In this Letter, we 
investigated systematic understanding for classes of the inflationary models 
from the viewpoint of the local conformal symmetry with a
slightly broken global symmetry in the metric-affine geometry. Contrary to the Riemannian geometry adopted in General Relativity, the metric-affine geometry regards both the metric and the affine connection as independent variables.
Consequently the Ricci curvature transforms covariantly and each term in the Lagrangian can preserve the local conformal invariance by itself, introducing the covariant derivative $D_\mu S=\pqty{\partial_\mu-\frac{1}{8}Q_\mu}S$ for a scalar $S$ with the non-metricity $Q_\mu=-g^{\alpha\beta}\nabla_\mu g_{\alpha\beta}$.
This allows much richer structures for theories with further local/global symmetries.
As simple examples we showed that the well-known $\alpha$-attractor and natural inflation can be systematically derived by the slightly broken $\SO(1,1)$ and $\groupO(2)$ global symmetries, respectively.

Our conclusions are not restricted to these two specific examples.
Noting the covariant transformations of curvature tensors $R_{\mu\nu}(\Gamma)\to R_{\mu\nu}(\Gamma)$ and $R(g,\Gamma)\to\ee^{2\sigma(x)}R(g,\Gamma)$ under the local conformal transformation $g_{\mu\nu}\to\ee^{-2\sigma(x)}g_{\mu\nu}$, one can construct a higher curvature gravity theory like Starobinsky's inflation~\cite{Starobinsky:1980te} in a conformal-invariant way (see, e.g., Ref.~\cite{Edery:2019txq}). There the quadratic term in the (anti-symmetric) Ricci tensor $R_{[\mu\nu]}=\pqty{R_{\mu\nu}-R_{\nu\mu}}/2$ 
gives rise to the kinetic term of the non-metricity $Q_\mu$. 
Therefore the non-metricity $Q_\mu$ can be a dynamical degree of freedom known as the \emph{Weyl gauge field}
rather than an auxiliary field~\cite{Ghilencea:2018dqd,Ghilencea:2019jux}. However it could have a Planck-scale mass and be irrelevant to inflation and low energy physics~\cite{Ghilencea:2020piz,Ghilencea:2020rxc}.
Also, once the scalar kinetic term is written in the covariant derivative as $X=-D_\mu SD^\mu S/2$,
the kinetic term transforms covariantly
under the local conformal transformation.
Therefore, one can easily construct
a non-canonical kinetic-term theory $\calL=P(X,S)$ with the local conformal invariance as a generalisation of $k$-inflation~\cite{ArmendarizPicon:1999rj}.
Multi-scalar generalisation is also a possible extension~\cite{Kallosh:2013daa, Azri:2019ffj}.
For a construction of the realistic inflationary model,
the completion of the reheating should be important, and hence
it should be interesting to consider the coupling to the matter in our framework~\cite{Brax:2014baa,Cespedes:2015jga,Bernardo:2017xcm}. We leave such
interesting
possibilities for future issues.

\acknowledgments

Y.T. is supported by JSPS KAKENHI Grants 
No. JP18J01992 and No. JP19K14707.
S.Y. is supported by JSPS Grant-in-Aid for Scientific Research(B) No. JP20H01932 and (C) No. JP20K03968.

\bibliography{main}
\bibliographystyle{JHEP.bst}
\end{document}